\documentclass[twocolumn,nofootinbib,prd,aps,tightenlines]{revtex4}

\usepackage{slashed}
\usepackage{braket}
\usepackage{graphicx}

{\count255=\time\divide\count255 by 60 \xdef\hourmin{\number\count255}
  \multiply\count255 by-60\advance\count255 by\time
  \xdef\hourmin{\hourmin:\ifnum\count255<10 0\fi\the\count255}}

\newcommand{\nn}{\nonumber \\ }

\def\slog{\mathsf{L}}

\def\lQ{\mathsf{L}_Q}
\def\lM{\mathsf{L}_M}
\def\darr#1{\raise1.5ex\hbox{$\leftrightarrow$}\mkern-16.5mu #1}

\begin{document}

\title{Electroweak Sudakov Corrections using Effective Field Theory}

\author{Jui-yu Chiu}
\affiliation{Department of Physics, University of California at San Diego,
  La Jolla, CA 92093}

\author{Frank Golf}
\affiliation{Department of Physics, University of California at San Diego,
  La Jolla, CA 92093}

\author{Randall Kelley}
\affiliation{Department of Physics, University of California at San Diego,
  La Jolla, CA 92093}

\author{Aneesh V.~Manohar}
\affiliation{Department of Physics, University of California at San Diego,
  La Jolla, CA 92093}

\begin{abstract}

Electroweak Sudakov corrections of the form $\alpha^n \log^m {s/M_{W,Z}^2}$ are summed using renormalization group evolution in soft-collinear effective theory (SCET). Results are given for the scalar, vector and tensor form-factors for fermion and scalar particles. The formalism for including massive gauge bosons in SCET is developed.
\end{abstract}

\date{\today\quad\hourmin}

\maketitle

The Large Hadron Collider (LHC) will study processes at energies $\sqrt{s}$ much larger than the masses of the $W$ and $Z$ bosons. Radiative (Sudakov) corrections to such processes contain logarithms $\slog=\log(s/M_{W,Z}^2)$, and one gets two factors of $\slog$ for each order in perturbation theory~\cite{ccc}. At LHC energies, these Sudakov corrections are not small --- a typical electroweak correction $\alpha \slog^2/(4 \pi \sin^2 \theta_W) \sim 0.15$ at $\sqrt{s}=4$~TeV. It is important to include these radiative corrections, since many new physics searches at the LHC look for small deviations from the standard model rate in high energy processes.

Electroweak Sudakov effects have been extensively studied recently~\cite{ciafaloni,fadin,kps,fkps,jkps,jkps4,beccaria,dp1,dp2,hori,beenakker,dmp,pozzorini,js}. Previous methods have relied on infrared evolution equations~\cite{fadin}, based on an analysis of the infrared structure of the perturbation theory amplitude and a factorization theorem for the Sudakov form factor~\cite{pqcd}. These summations have been  checked against one-loop~\cite{beccaria,dp1,dp2} and two-loop~\cite{hori,beenakker,dmp,pozzorini,js} computations. 

In this paper, electroweak Sudakov corrections will be summed using SCET~\cite{BFL,SCET}. An advantage of the SCET approach is that it divides the full computation into several simpler pieces, each of which involves a single scale---the matching corrections at $s$ and $M_{W,Z}^2$, and the operator anomalous dimension. This allows one to easily identify which quantities are universal, and which ones depend on the specific process, and to extend the results to new processes. The physical quantity we study is the Sudakov form factor---the amplitude $F(s)=\braket{p_1,p_2| \mathcal{O} | 0}$ for two on-shell particles  to be produced from the vacuum by an operator $\mathcal{O}$, with $s=(p_1+p_2)^2$. $F(s)$ for the fermion vector current $\bar \psi \gamma^\mu \psi$ for massless fermions has been computed previously, and so allows us to check the SCET results against other methods. We also compute $F(s)$ for a general fermion bilinear $\bar \psi \Gamma \psi$, for scalar operators $\phi^\dagger \phi$ and $i \phi^\dagger \darr{D}^\mu \phi$, and  for $\bar \psi \phi$, which are new results. The extension to massive fermions and scalars, including Higgs exchange corrections from the Yukawa couplings, is given in a longer article~\cite{cgkm}. 

The Sudakov form factor involves two energetic particles. LHC processes, such as 2-jet production, top pair production or squark production, typically involve four or more energetic particles. With our calculation method, it is easy to extend the results given here to these processes measured at the LHC without further computations---the anomalous dimensions and matching conditions in the effective theory are given by summing the results for the Sudakov form factor over all pairs of particles, and correcting for wavefunction renormalization~\cite{cgkm}.

We use the theory of Ref.~\cite{js}, a $SU(2)$ spontaneously broken gauge theory with a Higgs in the fundamental, where all gauge bosons get a common mass $M$. It is convenient, as in Ref.~\cite{js}, to write the group theory factors using $C_F$, $C_A$, $T_F$ and $n_F$, where $2n_F$ is the number of weak doublets.\footnote{Note that the results only hold for $C_A=2$, since for an $SU(N)$ group with $N>2$, a fundamental Higgs does not break the gauge symmetry completely.} We discuss this theory in the bulk of the paper, and  show how the results are modified for the $SU(3)\times SU(2)\times U(1)$ gauge theory of the standard model at the end. One important difference between the electroweak theory and QCD is that the gauge boson is massive, and we explain the formalism needed to include massive gauge bosons in SCET. We use the notation $a(\mu)=\alpha(\mu) /(4 \pi)$, $\lQ=\log Q^2/\mu^2$, $\lM=\log M^2/\mu^2$.

We will compute the Sudakov form factor $F_E(Q^2)$ in the Euclidean region for the spacelike process $\braket{p_2|\mathcal{O}|p_1}$ with $Q^2=-(p_2-p_1)^2 >0$ to avoid branch cuts in the Feynman integrals. The calculation will follow the discussion of deep-inelastic scattering as $x \to 1$  in Ref.~\cite{dis} (see also Ref.~\cite{ira}). The timelike Sudakov form factor is then given by analytic continuation, $F(s)=F_E(-s-i0^+)$, so that $\log (Q^2/\mu^2) \to \log(s/\mu^2)-i\pi$.

\begin{figure}
\begin{center}
\includegraphics[width=2.6cm]{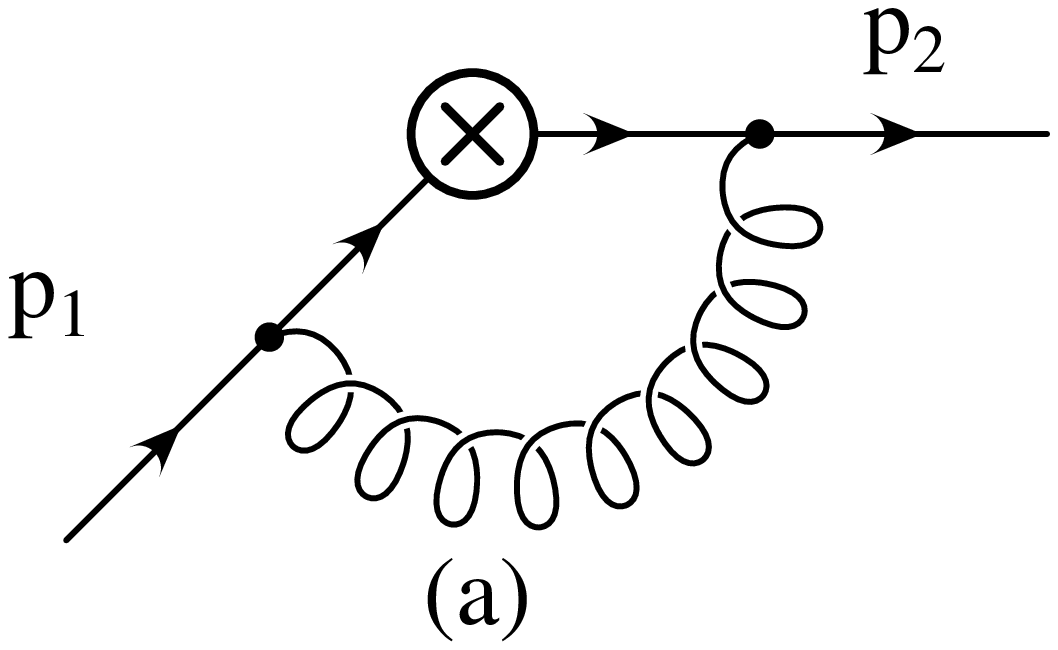}\qquad\includegraphics[width=2.6cm]{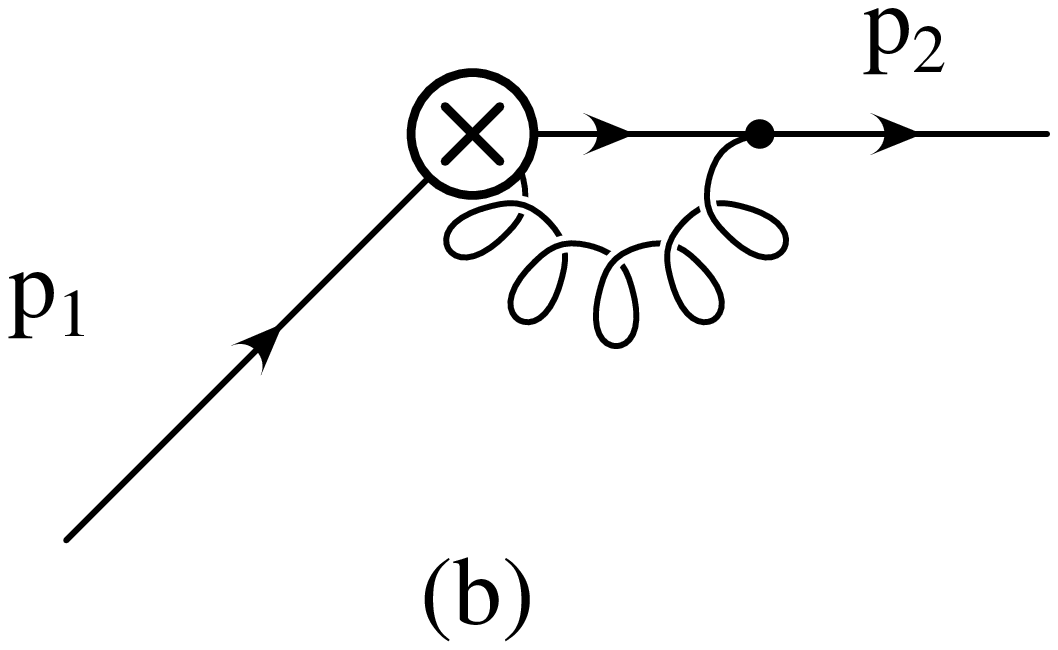}
\end{center}
\caption{\label{fig:match} Graphs contributing to the matching condition $C(\alpha(Q))$.
The solid line can be either a fermion or scalar.
The second graph only exists for the scalar case $\mathcal{O}=i \phi^\dagger \darr{D}^\mu \phi$. }
\end{figure}

The first step in the SCET computation is to match from the operator $\mathcal{O}$ in the full theory to the operator $\widetilde\mathcal{O}$ in SCET at the scale $\mu \sim Q$,
\begin{eqnarray}
\bar \psi \Gamma \psi &\to& \exp C(\mu)\, [\bar \xi_{n,p_2} W_n] \Gamma
[W^\dagger_{\bar n} \xi_{\bar n,p_1}] \nn
\phi^\dagger \phi &\to& \exp C(\mu)\, [ \Phi_{n,p_2}^\dagger W_n] 
[W^\dagger_{\bar n} \Phi_{\bar n,p_1}] \nn
i \phi^\dagger \darr{D}^\mu \phi &\to& \exp C(\mu)\, [ \Phi_{n,p_2}^\dagger W_n] i(\mathcal{D}_1+\mathcal{D}_2)^\mu[W^\dagger_{\bar n} \Phi_{\bar n,p_1}] \nn
\bar \psi \phi &\to& \exp C(\mu)\, [\bar \xi_{n,p_2} W_n] 
[W^\dagger_{\bar n} \Phi_{\bar n,p_1}]
\label{1}
\end{eqnarray}
where $i\mathcal{D}_1 = p_1+g( n \cdot A_{\bar n, q}) \frac{\bar n}{2}$,  $i\mathcal{D}_2 = p_2+g(\bar n \cdot A_{n, -q}) \frac{n}{2}$, $\xi,\Phi, A$ are the SCET fermion, scalar, and gauge fields, and $C(\mu)$ depends on the operator being matched. We have written the matching coefficient as $\exp C$ rather than $C$ for later convenience. The $n$-collinear direction is defined to be along $p_2$, and the $\bar n$-collinear direction along $p_1$, with $n=(1,0,0,1)$ and $\bar n=(1,0,0,-1)$. The light-cone components of a four-vector $p$ are defined by $p^+ \equiv n \cdot p$, $p^- \equiv \bar n \cdot p$. As is well-known, the matching coefficient can be computed as the finite part of the full theory graph, evaluated on-shell, with all infrared scales, such as the gauge boson mass set to zero (see e.g.~\cite{dis,hqet}).
The graphs to be evaluated are those in Fig.~\ref{fig:match}.
and the wavefunction graphs. The computation for $\mathcal{O}=\bar\psi \gamma^\mu \psi$ is identical to that for DIS~\cite{dis}, since the gauge boson mass is an infrared scale, and can be set to zero in the matching computation. The one-loop values of $C(\mu)$ for the other cases are computed similarly, and are given in Table~\ref{tab:results}, where $C(\mu)=C^{(1)}\alpha(\mu)/(4\pi)$ defines the one-loop correction $C^{(1)}$. The matching coefficient at the high scale will be chosen to be $C(\mu=Q)$, and is given by the second column in Table~\ref{tab:results} with $\lQ\to0$. There are no large logarithms in this matching correction.
\begin{table}
\begin{eqnarray*}
\begin{array}{|c|c|c|c|}
\hline
\mathcal{O} & C^{(1)}(\mu)/C_F & \gamma^{(1)}(\mu)/C_F & D^{(1)}(\mu)/C_F \\
\hline
\bar \psi  \psi &  - \lQ^2 +\frac{\pi^2}{6}-2&4 \lQ- 6 &\mathsf{X}-3\lM+\frac92\\[5pt]
\bar \psi \gamma^\mu \psi & -\lQ^2 +3\lQ+\frac{\pi^2}{6}-8&  4 \lQ- 6& \mathsf{X}-3\lM+\frac92 \\[5pt]
\bar \psi \sigma^{\mu\nu} \psi & - \lQ^2+4\lQ+\frac{\pi^2}{6}-8 & 4 \lQ- 6&\mathsf{X}-3\lM+\frac92 \\[5pt]
\phi^\dagger \phi &- \lQ^2+\lQ+\frac{\pi^2}{6}-2 &4 \lQ- 8 &\mathsf{X}-4\lM+\frac{7}{2} \\[5pt]
i \phi^\dagger \darr{D}^\mu \phi  & - \lQ^2 +4\lQ+\frac{\pi^2}{6}-8&  4 \lQ- 8 &\mathsf{X}-4\lM+\frac{7}{2}\\[5pt]
\bar \psi \phi & - \lQ^2+2 \lQ+\frac{\pi^2}{6}-4&  4 \lQ - 7&\mathsf{X}-\frac72\lM+4 \\[5pt]
\hline
\end{array}
\end{eqnarray*}
\caption{\label{tab:results} One-loop corrections to the Sudakov form-factor. $C^{(1)}$, $\gamma^{(1)}$ and $D^{(1)}$ are the coefficients of $a(\mu)$, and $\mathsf{X}\equiv-\lM^2+2\lM\lQ-5\pi^2/6$.}
\end{table}

\begin{figure}
\begin{center}
\includegraphics[width=2.6cm]{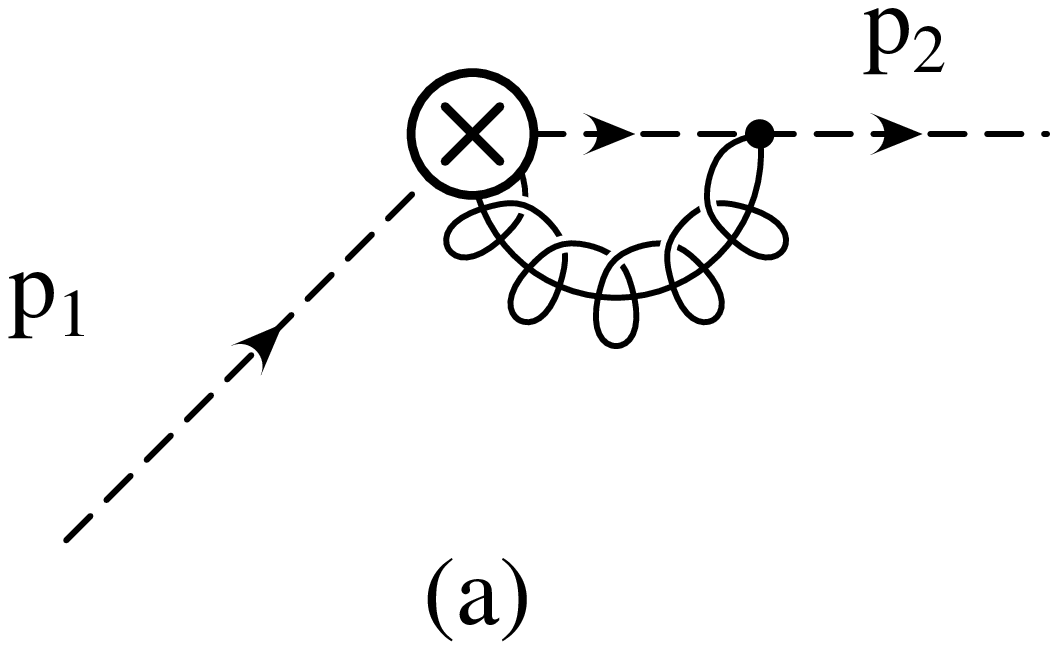}\quad\includegraphics[width=2.6cm]{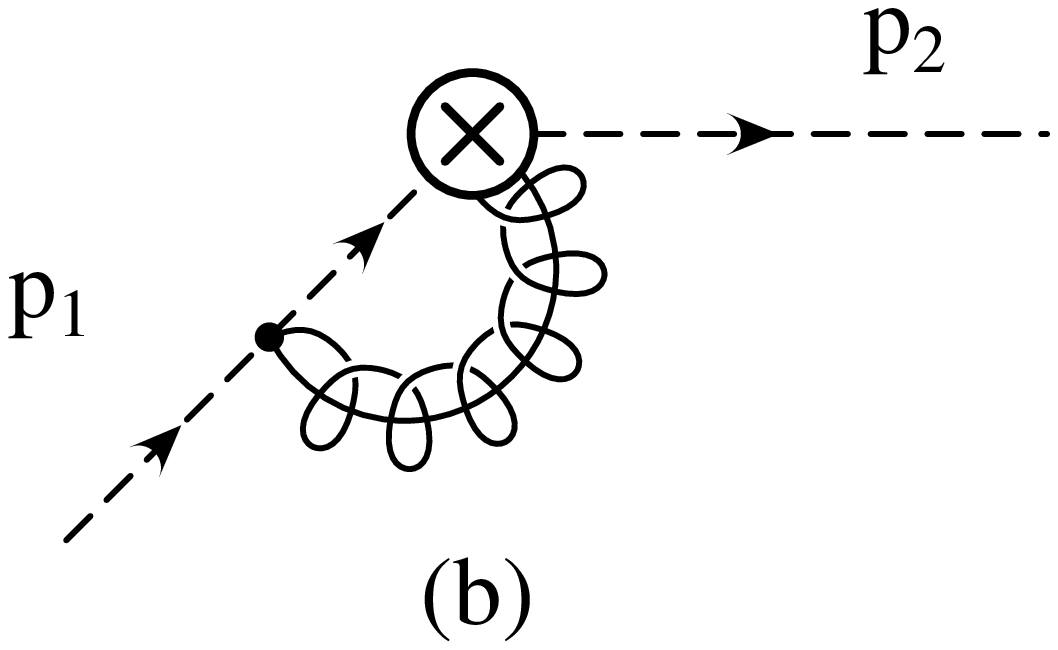}
\quad\includegraphics[width=2.6cm]{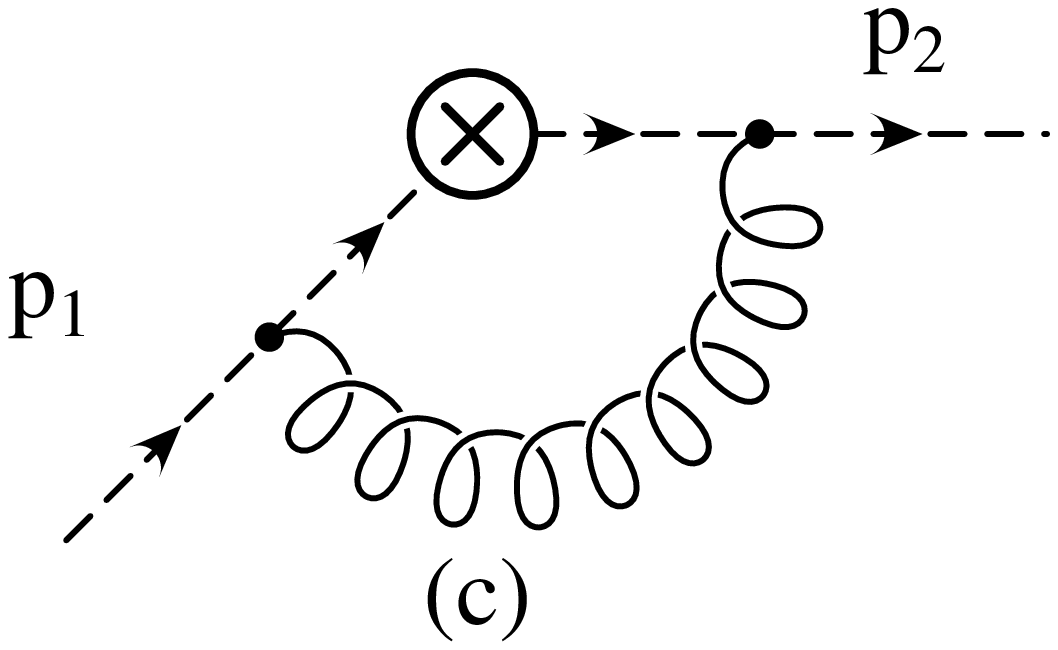}
\end{center}
\caption{\label{fig:scet} SCET (a) $n$-collinear, (b) $\bar n$-collinear and (c) ultrasoft graphs for the matrix element of $\widetilde\mathcal{O}$. The dotted lines are SCET propagators, and represent either fermions or scalars. There are also wavefunction graphs.}
\end{figure}

The renormalization group evolution of $\exp C(\mu)$ is given by the anomalous dimension of $\widetilde\mathcal{O}$ in SCET. The anomalous dimension is given by the ultraviolet counterterms for the SCET graphs in Fig.~\ref{fig:scet} (after zero-bin subtraction, see Ref~\cite{zerobin}). The ultraviolet divergence does not depend on the infrared properties of the theory, such as  a gauge boson mass, and for $\mathcal{O}=\bar \psi \gamma^\mu\psi $ is identical to the DIS result~\cite{dis}. The computations for the other cases is similar, and the results are given in Table~\ref{tab:results}, where $\mu \,{\rm d}C/{\rm d}\mu= \gamma_1$. The anomalous dimension $\gamma_1$ is used to evolve $C(\mu)$ from $\mu=Q$ down to the low scale $\mu=M$. The SCET anomalous dimension is linear in $\lQ=\log Q^2/\mu^2$~\cite{BFL}, and this form persists to all orders in perturbation theory~\cite{dis,Bauer:2003pi}, so we will write $\gamma_1(\mu) = A_1(\alpha(\mu) ) \lQ + B_1(\alpha(\mu))$. We will denote the one-loop corrections by $\gamma_1 = \gamma_1^{(1)} a$,  $A_1 = A_1^{(1)} a$ and $B_1 = B_1^{(1)} a$.

The final step in the computation is the matching condition at the low scale $\mu=M$. At this scale, the massive gauge boson is integrated out, and one matches to an effective theory which is SCET without the massive gauge boson. In our toy example, this effective theory contains no gauge particles. In the standard model, the effective theory has photons and gluons, but no $W$ and $Z$ bosons. The matching at $\mu=M$ is given by evaluating the graphs in Fig.~\ref{fig:scet}, and the wavefunction graphs. This computation is discussed in detail for $\mathcal{O}=\bar\psi \gamma^\mu\psi$, since it involves new features not discussed earlier in the literature. 
 
The $n$-collinear graph gives (omitting the factor $C(\mu)$)
\begin{eqnarray}
I_n &=&-i g^2 \mu^{2\epsilon} C_F  \int { {\rm d}^d k \over (2 \pi )^d} \nn
&&{\slashed{\bar n}\ n^\alpha \over 2}  {\slashed{n}\ \bar n \cdot (p_2-k)  \over 2 (p_2 - k )^2} \gamma^\mu { 1 \over  - \bar n \cdot k} \bar n_\alpha {1 \over k^2-M^2} \nn
&=& -2 i g^2 \mu^{2\epsilon}C_F \int { {\rm d}^d k \over (2 \pi )^d} 
  {\bar n \cdot (p_2-k)  \over (p_2 - k )^2} \gamma^\mu
{ 1 \over - \bar n \cdot k}  {1 \over k^2-M^2} .\nn
\label{2}
\end{eqnarray} 
This integral is divergent, even in $4-2\epsilon$ dimensions with an off-shellness, and needs to be regulated. We will regulate the integral by analytically continuing the fermion propagators, using an extension of the method given in Ref.~\cite{analytic}. The $p_i$ propagator denominator $(p_i-k)^2$ in the full theory  is analytically continued to
\begin{eqnarray}
\frac{1}{(p_i-k)^2} &\to& \frac{ (-\nu_i^2)^{\delta_i}}
{\left[(p_i-k)^2\right]^{1+\delta_i}}\: .
\label{3}
\end{eqnarray}
where $\nu_i$ and $\delta_i$ are new parameters. The $(p_2-k)^2$ denominator in Eq.~(\ref{2}) arises from the collinear $p_2$ propagator, and so gets modified as in Eq.~(\ref{3}). The $-\bar n \cdot k$ propagator in Eq.~(\ref{2}) arises from the $(p_1-k)^2$ propgator when $k$ becomes $n$-collinear. In this limit
\begin{eqnarray}
\frac{1}{(p_1-k)^2} &\to& \frac{ (-\nu_1^2)^{\delta_1}}
{\left[(n \cdot p_1) (-\bar n \cdot k)\right]^{1+\delta_1}}\: .
\label{4}
\end{eqnarray}
We will therefore analytically continue the $-\bar n \cdot k$ propagator in Eq.~(\ref{2}), which arises from the $W_n$ Wilson line~\cite{SCET} in $\mathcal{O}$ using
\begin{eqnarray}
\frac{1}{-\bar n \cdot k} &\to& \frac{(-\nu_1^-)^{\delta_1}}{(-\bar n \cdot k)^{1+\delta_1}}
\label{5}
\end{eqnarray}
where $\nu_1^- \equiv \nu_1^2/ p_1^+$. With this choice, Eq.~(\ref{2}) gives
\begin{eqnarray}
I_n   &=&  -2{\alpha \over 4 \pi}C_F  \gamma^\mu \left(\frac{\mu^2 e^{\gamma_E}}{M^2}\right)^{\epsilon} \left(\frac{\nu_2^2}{M^2}\right)^{\delta_2}  \left( \frac{\nu_1^-}{p_2^-}\right)^{\delta_1}      \nn
&& \times \frac{\Gamma(\epsilon+\delta_2)}{\Gamma(1+\delta_2)} \frac{\Gamma(2-\epsilon-\delta_2)\Gamma(\delta_2-\delta_1)}{\Gamma(2-\epsilon-\delta_1)}
\label{6}
\end{eqnarray}
The regulated value of $I_n$ is given by setting $\delta_i = r_i \delta$ and taking the limit $\delta \to 0$ first, followed by $\epsilon \to 0$~\cite{analytic},
\begin{eqnarray}
I_n&=&a C_F \gamma^\mu \Biggl[ \frac{2}{r_1-r_2}\frac{1}{\delta \epsilon}
+\frac{2}{r_1-r_2} \frac{1}{\delta} \log \frac{\mu^2}{M^2}-\frac{2r_2}{r_1-r_2}\frac{1}{\epsilon^2}\nn
&& \frac{1}{\epsilon}\biggl(2 + \frac{2r_1}{r_1-r_2} \log \frac{\nu_1^-}{p_2^-}
+\frac{2r_2}{r_1-r_2} \log \frac{\nu_2^2}{\mu^2} \biggr) \nn
 &&+2+2 \log \frac{\mu^2}{M^2}+\frac{2r_2}{r_1-r_2} \log \frac{\mu^2}{M^2}
 \log \frac{\nu_2^2}{\mu^2}\nn
 &&+\frac{2r_1}{r_1-r_2} \log \frac{\mu^2}{M^2}
 \log \frac{\nu_1^-}{p_2^-} + \frac{r_2}{r_1-r_2} \log^2 \frac{\mu^2}{M^2}\nn
&& +\frac{r_2 \pi^2}{2(r_1-r_2)}-\frac{r_1 \pi^2}{3(r_1-r_2)}
\Biggr],
\label{8}
\end{eqnarray}
which is a boost invariant expression, since $\nu_2^-/p_2^-$ is boost invariant.

The $\bar n$-collinear graph is given by Eq.~(\ref{6}) with the replacements $\delta_1 \leftrightarrow \delta_2$, $\nu_2 \to \nu_1$, $\nu_1^- \to \nu_2^+$, $p_2^- \to p_1^+$, with $\nu_2^+ \equiv \nu_2^2/p_2^-$. The parameters $\nu_2^+$ and $\nu_1^-$ play the same role as $\mu^\pm$ in the rapidity regularization method of Ref.~\cite{zerobin}.

The ultrasoft graph in Fig.~\ref{fig:scet} is regulated by the same method. The $p_2$ propagator $(p_2-k)^2$ is multipole expanded in the effective theory, and becomes
$-p_2^- k^+$, where $p_2^-$ is a label momentum (the $p_2$ subscript on $\xi_{n,p_2}$). Using Eq.~(\ref{3}) for the fermion propagators, we see that after multipole expansion, they are regulated in the same way as the Wilson line propagators. The ultrasoft graph gives
\begin{eqnarray}
I_s &=&-ig^2C_F \int { {\rm d}^d k \over (2 \pi )^d}n^\alpha {(-\nu_2^+)^{\delta_2}  \over [n \cdot (p_2-k)]^{1+\delta_2}} \nn
&&\times \gamma^\mu
{ (-\nu_1^-)^{\delta_1} \over [\bar n \cdot (p_1 -k )]^{1+\delta_1}}
\bar n_\alpha {1 \over k^2-M^2}
\label{7}
\end{eqnarray} 
and vanishes on-shell, since $p_2^+=p_1^-=0$. The total SCET contribution $I_n+I_{\bar n}+I_s$ plus the wavefunction  graphs is
\begin{eqnarray}
aC_F\left[ \frac{2}{\epsilon^2} +\frac{\left(3-2 \lQ\right)}{\epsilon}  -\lM^2 +2 \lM \lQ-3 \lM+\frac92  -\frac{5 \pi^2}{6} \right]\nn
\label{9}
\end{eqnarray}
where we have used $\nu_1^-=\nu_1^2/p_1^+$, $\nu_2^+=\nu_2^2/p_2^-$, and $Q^2=p_1^+ p_2^-$. The dependence on $r_{1,2},\delta$ and $\nu^2$ has dropped out. The $1/\epsilon$ poles are cancelled by the ultraviolet counterterms in the effective theory, and give the same anomalous dimension as in Table~\ref{tab:results}. The contributions of the various diagrams to the anomalous dimension  in our calculation is different from that in previous results using an off-shell regulator~\cite{dis}, where the ultrasoft graph is non-zero, and contributes to $\gamma_1$. The finite part of Eq.~(\ref{9}) gives the multiplicative matching correction $\exp D(\mu)$, $D(\mu)=D^{(1)}\alpha(\mu)/(4\pi)$ when the massive gauge boson is integrated out. The other cases are computed similarly, and are given in Table~\ref{tab:results}. The massive gauge boson can be integrated out at the scale $\mu=M$, so that $\lM \to 0$. In this case, there are no large logarithms in the matching, since $\lQ$ only occurs multiplied by $\lM$ in $D(\mu)$. This is an accident of the one-loop computation~\cite{cgkm}. We show in Ref.~\cite{cgkm} that in general, one can have a single power of $\lQ$ in the matching condition $D$ at higher order. This is consistent with Eq.~(\ref{9}), which has a single $\lQ$ term if $\mu$ is chosen to be of order, but not exactly equal to, $M$. 

The theory below $\mu=M$, SCET with the massive gauge boson integrated out, is a free theory in our example, so the operator matrix elements are given by their tree-level value.  There is no need to introduce any propagating gauge modes below $M$~\cite{ir}. The one-loop renormalization group improved value for the Sudakov form factor is Eq.~(\ref{11}) with $\gamma_2\to0$.
This can be compared with fixed order results by expanding this in a power series expansion in $\alpha(M)$, and correctly reproduces the known $\alpha \slog$, $\alpha^2 \slog^4$ and $\alpha^2 \slog^3$ terms. Including, in addition, the known two-loop cusp anomalous dimension~\cite{cusp}, which gives the two-loop value for $A_1$ reproduces the $\alpha^2 \slog^2$ term. The $\alpha^2 \slog$ term requires the two-loop $B_1$ term in $\gamma_1$. Including the two-loop cusp anomalous dimension sums the LL and NLL Sudakov series. The two-loop value for $B_1$ involves graphs with Higgs loops, and is not known.

The results can be extended to the Sudakov form factor in the standard model. The form-factor has to be computed separately for $\bar \psi \gamma^\mu P_{L,R} \psi$ since the theory is chiral. The operators $\bar \psi P_{L,R} \psi$ and $\bar \psi \sigma^{\mu\nu} P_{L,R}\psi$ are not gauge invariant, so we do not consider them. We give the results in the limit of massless external particles, so that Higgs exchange contributions can be neglected. Flavor mixing and Higgs contributions can easily be included~\cite{cgkm}.

The matching at $Q$ and the anomalous dimension $\gamma_1$ for the other operators are given by replacing $\alpha C_F$ by $\sum_i \alpha_i C_{Fi}$, where $\alpha_3=\alpha_s$, $\alpha_2=\alpha/\sin^2 \theta_W$, $\alpha_1=\alpha/\cos^2\theta_W$, $C_{F3}$ and $C_{F2}$ are the color and weak Casimirs, and $C_{F1}$ is $Y^2$, where $Q=T_3+Y$. The form-factor is evolved down to a scale of order the gauge boson mass. For definiteness, we choose $\mu=M_Z$, at which point the $W$ and $Z$ are integrated out.\footnote{One can also integrate out the $Z$ and $W$ sequentially in two steps. This sums powers of $\log M_Z/M_W$, which is a large log in the $\cos \theta_W \to 0$ limit.} The multiplicative matching correction at $\mu=M_Z$ is given by adding $D$ with $\alpha C_F \to \alpha (T_3-\sin^2\theta_W Q)^2/(\sin^2 \theta_W \cos^2\theta_W)$ and $M=M_Z$ to $D$ with $\alpha C_F \to  \alpha (T^2-T_3^2)/\sin^2\theta_W $ and $M=M_W$. Below this scale, the theory reduces to a gauge theory with gluons and photons, so the operator coefficient has an anomalous dimension $\gamma_2(\mu)$ equal to $\gamma_1$ with the replacement $\alpha C_F \to \alpha_s C_{F3} + \alpha Q^2$. The final expression for the form-factor in the standard model is then
\begin{eqnarray}
&&\log F_E(Q^2,\mu) = C(\mu=Q) + \int_{Q}^{M_Z} \frac{ {\rm d}\mu}{\mu}
\gamma_1(\mu)\nn
&& + D_{Z,W}(\mu=M_Z) +  \int_{M_Z}^{\mu} \frac{ {\rm d}\mu}{\mu}
\gamma_2(\mu) . 
\label{11}
\end{eqnarray}
This equation is used to evolve $\mathcal{O}$ down to some low-scale, which depends on the physical process being considered. The $\mu$ dependence cancels between the operator anomalous dimension, and the matrix element computed in the effective theory. For example, if one is interested in the cross-section for two-jet events, $\mu$ can be chosen to be the jet invariant mass, and the cross-section computed in SCET~\cite{2jet}.

Equation~(\ref{11}) can be used to compute the electroweak radiative corrections.
\begin{table}
\begin{eqnarray*}
\begin{array}{|c|c|c|c|c|c|c|c|c|c|}
\hline
 & \multicolumn{4}{c|}{\mu=M_Z} & \multicolumn{4}{c|}{\mu=30~\text{GeV}} \\
 \hline
Q\text{(TeV)}& 0.5 & 1 & 4 & 10 & 0.5 & 1 & 4 & 10\\
\hline\hline
e & 0.98 & 0.96 & 0.91 & 0.86 & 0.97 & 0.95 & 0.89 & 0.84\\
u & 0.90 & 0.82 & 0.59 &0.43 & 0.74 & 0.62 & 0.38 & 0.25 \\
\hline
\end{array}
\end{eqnarray*}
\caption{\label{tab:numerics}  $F_E(Q^2)$ for electron production via $\bar L \gamma^\mu P_L L$, and $u$-quark  production via $\bar Q \gamma^\mu P_L Q$, where $L$ and $Q$ are the lepton and quark doublets.}
\end{table}
Table~\ref{tab:numerics} gives the numerical values of $F_E(Q^2)$ for a few sample values of parameters, for scaling down to $\mu=M_Z$, and to $\mu=30$~GeV, the typical invariant mass used to define a jet at the LHC.  The numerical values are slightly smaller than the estimate in the introduction, because of cancellation between the two terms in the anomalous dimension $\propto 4\lQ-6$. 

The Sudakov form factor was considered here because it provided
a simple example of our method, with the effective theory operator involving only two external fields. The same methods can be applied to processes of direct relevance to the LHC, such as quark pair production, or the production of new particles such as squarks~\cite{cgkm}. These applications require operators involving four external fields. The dominant part of the anomalous dimension (the $\lQ$ term) for four-particle operators is twice that for the two-particle operators, so the radiative corrections for pair production are about twice as large as those computed here. The effective theory method readily generalizes to particle production, and to other applications. In particular, it is possible to include $SU(2) \times U(1)$ mixing effects (i.e.\ $M_W \not = M_Z$) as show here, and to include Higgs radiative corrections, which depend on the large $t$-quark Yukawa coupling. A more extensive discussion is given in a longer publication~\cite{cgkm}.

RK was supported by an LHC theory fellowship from the NSF.


\begin{thebibliography}{99}

\bibitem{ccc}
  M.~Ciafaloni, P.~Ciafaloni and D.~Comelli,
  Phys.\ Rev.\ Lett.\  {\bf 84}, 4810 (2000)

\bibitem{ciafaloni}
  P.~Ciafaloni and D.~Comelli,
  Phys.\ Lett.\  B {\bf 446}, 278 (1999);
  Phys.\ Lett.\  B {\bf 476}, 49 (2000)

\bibitem{fadin}
  V.~S.~Fadin, L.~N.~Lipatov, A.~D.~Martin and M.~Melles,
  Phys.\ Rev.\  D {\bf 61}, 094002 (2000)
  
\bibitem{kps}
  J.~H.~Kuhn, A.~A.~Penin and V.~A.~Smirnov,
  Eur.\ Phys.\ J.\  C {\bf 17}, 97 (2000)
  
  
\bibitem{fkps}
  B.~Feucht, J.~H.~Kuhn, A.~A.~Penin and V.~A.~Smirnov,
  Phys.\ Rev.\ Lett.\  {\bf 93}, 101802 (2004)
  
  
\bibitem{jkps}
  B.~Jantzen, J.~H.~Kuhn, A.~A.~Penin and V.~A.~Smirnov,
  Phys.\ Rev.\  D {\bf 72}, 051301 (2005)
  [Erratum-ibid.\  D {\bf 74}, 019901 (2006)]
  
\bibitem{jkps4}
  B.~Jantzen, J.~H.~Kuhn, A.~A.~Penin and V.~A.~Smirnov,
  Nucl.\ Phys.\  B {\bf 731}, 188 (2005)
  [Erratum-ibid.\  B {\bf 752}, 327 (2006)]
  

\bibitem{beccaria}
  M.~Beccaria, F.~M.~Renard and C.~Verzegnassi,
  Phys.\ Rev.\  D {\bf 63}, 053013 (2001)

\bibitem{dp1}
  A.~Denner and S.~Pozzorini,
  Eur.\ Phys.\ J.\  C {\bf 18}, 461 (2001)

\bibitem{dp2}
  A.~Denner and S.~Pozzorini,
  Eur.\ Phys.\ J.\  C {\bf 21}, 63 (2001)
 
\bibitem{hori}
  M.~Hori, H.~Kawamura and J.~Kodaira,
  Phys.\ Lett.\  B {\bf 491}, 275 (2000)

\bibitem{beenakker}
  W.~Beenakker and A.~Werthenbach,
  Nucl.\ Phys.\  B {\bf 630}, 3 (2002)

\bibitem{dmp}
  A.~Denner, M.~Melles and S.~Pozzorini,
  Nucl.\ Phys.\  B {\bf 662}, 299 (2003)
  
\bibitem{pozzorini}
  S.~Pozzorini,
  Nucl.\ Phys.\  B {\bf 692}, 135 (2004)
  
\bibitem{js}
  B.~Jantzen and V.~A.~Smirnov,
  Eur.\ Phys.\ J.\  C {\bf 47}, 671 (2006)

\bibitem{pqcd}
 A.~H.~Mueller, {\it Perturbative Quantum Chromodynamics},
 World Scientific, Singapore, 1989.
  
  \bibitem{BFL}
C.~W.~Bauer, S.~Fleming and M.~E.~Luke,
Phys.\ Rev.\ D {\bf 63}, 014006 (2001).

\bibitem{SCET}
C.~W.~Bauer, S.~Fleming, D.~Pirjol and I.~W.~Stewart,
Phys.\ Rev.\ D {\bf 63}, 114020 (2001);
C.~W.~Bauer and I.~W.~Stewart,
Phys.\ Lett.\ B {\bf 516}, 134 (2001);
C.~W.~Bauer, D.~Pirjol and I.~W.~Stewart,
Phys.\ Rev.\ D {\bf 65}, 054022 (2002).

  
 \bibitem{cgkm}
  J.~y.~Chiu, F.~Golf, R.~Kelley and A.~V.~Manohar,
  arXiv:0712.0396 [hep-ph].
 
\bibitem{dis}
  A.~V.~Manohar,
  Phys.\ Rev.\  D {\bf 68}, 114019 (2003)

\bibitem{ira}
  C.~W.~Bauer, S.~Fleming, D.~Pirjol, I.~Z.~Rothstein and I.~W.~Stewart,
  Phys.\ Rev.\  D {\bf 66}, 014017 (2002)

\bibitem{hqet}
  A.~V.~Manohar,
  Phys.\ Rev.\  D {\bf 56}, 230 (1997);
  arXiv:hep-ph/9606222.

\bibitem{zerobin}
  A.~V.~Manohar and I.~W.~Stewart,
  arXiv:hep-ph/0605001.

  

\bibitem{Bauer:2003pi}
  C.~W.~Bauer and A.~V.~Manohar,
  Phys.\ Rev.\  D {\bf 70}, 034024 (2004)

\bibitem{analytic}
  V.~A.~Smirnov and E.~R.~Rakhmetov,
  Theor.\ Math.\ Phys.\  {\bf 120}, 870 (1999)
  [Teor.\ Mat.\ Fiz.\  {\bf 120}, 64 (1999)]
  M.~Beneke and T.~Feldmann,
  Nucl.\ Phys.\  B {\bf 685}, 249 (2004)

\bibitem{ir}
  A.~V.~Manohar,
  Phys.\ Lett.\  B {\bf 633}, 729 (2006)
\bibitem{cusp}
  G.~P.~Korchemsky and A.~V.~Radyushkin,
  Nucl.\ Phys.\  B {\bf 283}, 342 (1987).

\bibitem{2jet}

  C.~W.~Bauer, C.~Lee, A.~V.~Manohar and M.~B.~Wise,
  Phys.\ Rev.\  D {\bf 70}, 034014 (2004)
  C.~W.~Bauer, A.~V.~Manohar and M.~B.~Wise,
  Phys.\ Rev.\ Lett.\  {\bf 91}, 122001 (2003)
  C.~W.~Bauer and M.~D.~Schwartz,
  Phys.\ Rev.\ Lett.\  {\bf 97}, 142001 (2006)
  M.~Trott,
  Phys.\ Rev.\  D {\bf 75}, 054011 (2007)

\end{thebibliography}
\end{document}